\documentclass[12pt,preprint]{aastex}

\slugcomment{Submitted to aastex} \shorttitle{Emission
lines,continuous injection and achromatic bump}
\shortauthors{Gao et al.}

\begin {document}

\title{CAN THE BUMP BE OBSERVED IN THE EARLY AFTERGLOW OF GRBS WITH X-RAY LINE EMISSION FEATURES?}

\author{W.H.Gao and D.M.Wei}
\affil{Purple Mountain Observatory,Chinese Academy of
Sciences,Nanjing,210008,China.} \affil{National Astronomical
Observatories,Chinese Academy of Sciences,China.}
\email{whgao@pmo.ac.cn, dmwei@pmo.ac.cn}
\begin{abstract}

    Extremely powerful emission lines are observed in the
  X-ray afterglow of several GRBs. The energy
  contained in the illuminating continuum which is
  responsible for the line production exceeds 10$^{51}$ erg,
  much higher than that of the collimated GRBs. It constrains
  the models which explain the production of X-ray emission lines. In
  this paper, We argue that this energy can come from
  a continuous postburst outflow. Focusing on a central engine of
  highly magnetized millisecond pulsar or magnetar
  we find that afterglow can be affected by the illuminating
  continuum, and therefore a distinct achromatic bump may be
  observed in the early afterglow lightcurves. With the luminosity of the
  continuous outflow which produces the line emission, we
  define the upper limit of the time when the bump feature
  appears.

    We argue that the reason why the achromatic bumps have not been
  detected so far is that the bumps should appear at the time too early to be observed.

\end{abstract}


\section{Introduction}

    X-ray emission lines observed in the X-ray afterglow of GRBs
  are important clues for identifying the nature of the
  progenitors of the long (t$\geq$ 2s) GRBs. So far there are at
  least eight GRBs to show evidence for X-ray emission
  features in the early afterglow. GRB970508 (Piro et al.1999) and
  GRB000214 (Antonelli et al.2000) were detected by
  Bepposax, GRB970828 (Yoshida et al.2001) by ASCA, GRB991216
  (Piro et al.2000) and GRB020813 (Butler et al.2003) by Chandra,
  GRB011211 (Reeves et al.2002), GRB001025A (Watson et al.2002) and
  GRB030227 (Watson et al.2003) by XMM-Newton.

    Mainly there are two classes of models to
  explain the X-ray lines in the X-ray afterglow, Geometry Dominated
  Models (GD models)(eg. Lazzati et al.1999) and Engine Dominated Models
  (ED models)(eg. Rees \& M$\acute{e}$sz$\acute{a}$ros 2000,
  M$\acute{e}$sz$\acute{a}$ros \& Rees 2001). The main
  difference between these two classes of models lies on how to explain the
  observed duration of the line emission and the energy where the
  X-ray lines come from. The detailed properties of X-ray emission
  lines can be seen in the paper of Lazzati (2002). We note
  here that there is a strict constraint on the energy powering the
  luminous lines in the Geometry Dominated Models. In GD models, since
  the input energy is taken from GRB or early afterglow continuum, this
  energy should not be larger than that of the GRB explosion. However in
  the ED models the energy powering the emission lines is assumed to be provided
  by a continuous injecting inner engine at the end of the GRB emission
  instead of turning off abruptly (Rees \& M$\acute{e}$sz$\acute{a}$ros
  2000). So less constraint can be put on the energy in ED models.

    It has been pointed out that the emission lines can be used to put a firm
  lower limit on the energy of the illuminating continuum (Lazzati 2002,
  Ghisellini et al.2002). These lines last at least several hours. It
  implies that the energy is of the order of 10$^{49}$ erg, therefore
  the energy contained in the illuminating continuum must
  exceed 10$^{51}$ erg, beyond
  the energy of the GRBs corrected by the estimate of
  their degree of collimation (Frail et al.2001,Bloom et al.2003).
  Therefore, it is difficult to explain the production of the emission lines
  with the energy of GRBs or the early afterglow continuum.

    Rees \& M$\acute{e}$sz$\acute{a}$ros (2000) considered that an
  extended, possibly magnetically dominated wind from a GRB
  impacting the expanding envelope of a massive progenitor can
  account for the observed emission features. In this case, luminosity
  as high as 10$^{47}$ erg s$^{-1}$ can be
  produced 1 day after the original explosion and last (but
  decaying) for enough time to support so high energy of
  illuminating continuum. It is natural that the Engine Dominated
  Models can solve the contradiction between the energy obtained
  from X-ray emission lines and that of the collimated
  GRBs. But in ED models the afterglow can be
  affected by the continuous injection.

    In this paper we investigate the Engine Dominated
  Models, focusing on the central engine is a highly magnetized
  millisecond pulsar or magnetar,following the work
  of Dai \& Lu (1998a,1998b),who first considered continuous
  injection from a millisecond pulsar to interpret the afterglow
  light curves of some GRBs. We also reanalyze
  the energy limited by the emission features and compare it with
  that of the collimated GRBs (Frail et al.2001,Bloom et
  al.2003). The cosmological parameters will be set through this paper to
  $H_{0}$=65 km s$^{-1}$Mpc$^{-1}$, $\Omega_{m}$=0.3, $\Omega_{\Lambda}$=0.7
  .

\section{Reanalyze the energies of observed emission features}

    The main information obtained for the seven bursts detected so
  far is presented in Table 1. The quantities listed in the Table
  are described as following:

    Redshift: GRB000214 did not have a standard
  optical or radio afterglow and therefore there is no identification of
  the host galaxy. The value of the redshift of this GRB was
  inferred from the X-ray line, identified  as a iron 6.97 keV
  emission line.

    Line luminosity: The line luminosity listed in the Table 1 is
  derived under the assumption of isotropic emission. The collimation
  of the line photon will enhance the observed flux by a factor
  4$\pi$/$\Omega_{line}$ with respect to isotropic emission, which has been
  discussed by Ghisellini et al.(2002). They found that if electron
  scattering was important, the amplification of the line could be at
  most a factor of two.

    Time of the line emission: the start time t$_{s}$ corresponds
  to the beginning of the emergence in the X-ray observation in the afterglow,
  while the end time t$_{e}$ is defined by the time when the line is not detected
  any longer or the time of the end of the observation though
  the line can still be detected.

    ISM Density: There are only two GRBs (GRB970508 and
  GRB991216) whose interstellar medium density has been derived.
  For GRB970508, the ambient medium density was found to be $\sim$
  1 cm$^{-3}$ from long time of the radio observation
  (Frail et al.2000). The interstellar medium density was
  calculated to be 4.7 cm$^{-3}$ by modelling the broadband
  emission of the afterglow of GRB991216 (Panaitescu $\&$ Kumer 2001).

    Energy of the observed X-ray line: A line of constant flux comes from
  GRB000214 (Antonelli et al.2000). For GRB970508 and GRB011211
  the emission lines became undetectable before the end of the X-ray
  observation, while for GRB020813 the lines could also be detected until
  the observation ended. For GRB030227, the emission lines were
  detected only in the final segment of the observation, implying
  that the lines not only faded but also appeared at a significant time
  after the GRB. Since we can not know if it is true for other
  GRBs and can not find out the exact time at which the emission
  lines appear, we adopt two values of duration of the lines as
  Ghisellini et al.(2002). The first is assumed that the lines exist
  only for the time interval t$_{e}$-t$_{s}$ ( " short lived" line ), while
  the second is assumed that the lines remain constant in flux
  for the interval t$_{e}$-0 ( " long lived " line ). So two values of
  the total line energy can be obtained.

    Lower limit constraint on the total energy of the burst: Emission
  lines can be used to
  constrain the total energy of the burst. In fact, the line
  energy is only a fraction $\eta_{line}$ of the illuminating
  continuum which is in turn a fraction $\eta_{x}$ of the
  energy emitted in $\gamma$-ray during the prompt
  emission. Considering the collimating of the line photons, we take
  the amplification factor $4\pi/\Omega_{line}$ with respect to
  the isotropic case.

    In the reflection mechanism the efficiency of the production
  of line photon $\eta_{line}$ was computed numerically by Lazzati
  et al.(2002). The reprocessing efficiency for iron can not be
  larger than 2$\%$, while the combined light elements, for very
  small ionization parameters, can reprocess up to 5$\%$ of the
  continuum into soft X-ray narrow lines. $\eta_{x}\sim$ 0.05-0.1
  is the value commonly observed during the prompt emission of
  most bursts, which corresponds to a spectrum
  $F(\nu)\varpropto\nu^{0}$ up to 300 keV (Lazzati 2002). It implicitly
  assumes that $\eta_{x}$ is same in the collimated cone. The
  amplification of the line luminosity after line photons
  reprocessed by scattering off the funnel wall has been discussed
  by Ghisellini et al.(2002). No or weak collimation of the line
  photons is found.

    The total energy derived from the emission line
  can be written as:

\begin{equation}
      E_{\gamma} \geq 2\frac{E^{iso}_{line}}{\eta_{x}\eta_{line}}
                       \frac{\Omega_{line}}{4\pi}
                 = 500E^{iso}_{line}(\frac{0.1}{\eta_{x}})
                 (\frac{0.02}{\eta_{line}})(\frac{\Omega_{line}/4\pi}{0.5})
\end{equation}

    Applying the fiducial values discussed above for the
  efficiencies involved,we obtain the total energy listed in the
  Table 1.

    For GRB011211 only emission lines of elements lighter than iron
  can be seen (Reeves et al.2002). We used $\eta_{line}$=0.05 to
  derive the lower limit for the energy radiated in $\gamma$-rays,
  we then obtain $E_{\gamma}\gtrsim $5$\times$10$^{50}$ erg and
  $E_{\gamma}\gtrsim $4.4$\times$10$^{51}$ erg for short and long
  lived line respectively. We find that the lower limit of the
  energy is about a factor of 4 to 33 larger
  than the energy estimated by Frail et al.(2001)( $E_{\gamma}$=
  1.32$\times$10$^{50}$ erg corrected by $\theta_{j}$=3.6 degree).

    For GRB020813,only the exact value of flux of S line
  can be measured. We derive $E_{\gamma}\gtrsim$ 3$\times$10$^{51}$
  erg and $E_{\gamma}\gtrsim$ 6$\times$10$^{51}$ erg for short and
  long lived cases, a factor
  $\sim$8 to 16 above the estimate of Frail et al.(2001) for this
  burst.

    The lower limit value of energy for GRB030227 is
  $E_{\gamma}\gtrsim$
  2.4$\times$10$^{51}$ erg and $E_{\gamma}\gtrsim$
  2.1$\times$10$^{52}$ erg for short and
  long lived cases. We do not know the time when the break
  appeared in the afterglow lightcurves after the trigger. If we
  take the value of energy for GRB030227 to be 5$\times$10$^{50}$ erg (Frail et
  al.2001), we find this value is a factor of 5 to 40 smaller than the lower
  limit of energy derived from the line emission.

    The detailed discussion on GRB970508,GRB000214 and GRB991216 has also been made by
  Ghisellini et al.(2002).

    Compared with the corrected $\gamma$-ray energy of the
  GRBs (Frail et al.2001,Bloom et al.2003), a quite significant discrepancy
  of the energy can be seen.

    Since the GRB explosion energy depends on n$^{\frac{1}{4}}$, to
  explain the discrepancy of the energy, Ghisellini et
  al.(2002) assume that circum-burst density should be
  0.1$\times(\frac{E_{\gamma}}{E_{\gamma}^{F01}})^{4}$cm$^{-3}$, larger
  than the values derived from modelling broad-band afterglow light curves
  (Panaitescu \& Kumer 2001).

    We consider this energy is attributable to the source of the illuminating
  continuum which produces the line features and does not come from GRB prompt
  explosion. A continuous postburst
  outflow is taken into account for the extremely powerful emission lines.

\section{Continuous injection from a highly magnetized millisecond  pulsar}

    A decaying magnetar model was suggested to explain the line
  emission in the X-ray afterglow (Rees \& M$\acute{e}$sz$\acute{a}$ros
  2000). In this model the outflow after the GRB explosion
  continues at a diminishing rate for a longer time of hours to
  days,not as the typical GRB model in which its energy and mass
  outflow are either a delta or a top-hat function. The prolonged
  activity could arise either due to a spinning-down millisecond
  pulsar or to a highly magnetized torus around a black
  hole, which could produce a luminosity as high as
  $L_{m}\sim$ 10$^{47}$t$_{day}^{-1.3}$erg s$^{-1}$ even one day
  after the original explosion to produce
  the observed lines.

    Although the ED models can explain the production of the extremely
  powerful X-ray lines, it is important to stress that the continuous
  injection which is reprocessed in the expanding envelope of a massive
  progenitor to produce the line features can affect the early
  afterglow lightcurves. In this case,achromatic bump may be found
  in the lightcurves of the afterglow.

    Considering the central engine that emits an initial impulsive input
  energy E$_{imp}$ as well as a continuous luminosity which
  varies as a power law in the emission time, in which a self-similar
  blast wave forms at the late times (Blandford $\&$ McKee 1976), the
  differential energy conservation relation
  (Zhang $\&$ M$\acute{e}$sz$\acute{a}$ros 2001) is

\begin{equation}
dE/dT=L-k(E/T)
\end{equation}

  Here L=L$_{0}$(T/T$_{0}$)$^{q}$ is the intrinsic luminosity of
  the central engine. The integrated relation is

\begin{equation}
 E=\frac{L_{0}}{k+q+1}(\frac{T}{T_{0}})^{q}
 T+E_{imp}(\frac{T}{T_{0}})^{-k}, T>T_{0}
\end{equation}

  when 1+q+k$\neq$0 (Cohen $\&$ Piran 1999,
  Zhang $\&$ M$\acute{e}$sz$\acute{a}$ros 2001). Here E and T denote the energy
  and time measured in the observer frame, and q and k are
  dimensionless constants. T$_{0}$ is a characteristic
  timescale for the formation of a
  self-similar blast wave, which is roughly equal to the time for
  the external shock to start to decelerate, and E$_{imp}$ is constant
  that describes the impulsive energy input. The first term denotes
  the continuous luminosity injection, and second term takes into account radiative
  energy losses in the blast wave. A self-similar blast wave is assumed to
  exist only at T$>$T$_{0}$. Setting T=T$_{0}$, total energy at the
  beginning of the self-similar expansion is
  E$_{0}$=L$_{0}$T$_{0}$/(1+k+q)+E$_{imp}$. The first term is the
  energy from the continuous injection before the self-similar
  solution starts. The second term is the energy injected
  impulsively by the initial event. Note that
  1+k+q$>$0, otherwise, E$_{0}$ no longer has a clear physical
  meaning for the first term will be negative.

    The total energy E discussed above may be dominated either by
  the continuous injection term($\propto$T$^{(1+q)}$) or by the
  impulsive term($\propto$T$^{(-k)}$). A distinctive influence will
  appear on the GRB afterglow lightcurves if the continuous
  injection term dominates over the impulsive term after a
  critical time T$_{c}$. By equating the injection and impulsive
  term we can define the critical time T$_{c}$ ,

\begin{equation}
T_{c}=Max\{1,[(1+k+q)\frac{E_{imp}}{L_{0}T_{0}}]\}T_{0}
\end{equation}
  noticing that T$_{c}$$>$T$_{0}$ must be satisfied to ensure that
  a self-similar solution has already formed when the continuous
  injection dominates. It is only T$>$T$_{c}$ that the continuous
  injection would have a distinctive effect on the lightcurves of
  the afterglow.

    We focus the central engine on a highly magnetized millisecond
  pulsar or magnetar to satisfy that the continuous injection varies as a
  approximate power law (Dai \& Lu 1998a,1998b). The continuous injection
  into the fireball may be
  mainly due to electromagnetic dipolar emission. For an
  electromagnetic-loss-dominated case, the luminosity of dipole radiation is
\begin{eqnarray}
L_{ill} = L_{em,0}(\frac{1}{1+\frac{T}{T_{em}}})^{2}
\simeq \left\{\begin{array}{cc}L_{em,0},& T\ll T_{em}\\
L_{em,0}(\frac{T}{T_{em}})^{-2},&T\gg T_{em}
\end{array}
\right.\end{eqnarray}

    where L$_{em,0}$ is the initial luminosity of the dipolar
  spin-down emission. In this case, L$_{0}$=L$_{em,0}$. T$_{em}$
  is the characteristic timescale for
  dipolar spin-down and T can also be defined as the time when the
  emission lines begin to appear, accordingly L$_{ill }$ is the
  luminosity of the X-ray illuminating continuum which produces the line
  features just at the time T. L$_{ill}$ can also be defined as following:

\begin{equation}
L_{ill} = \frac{L_{line}}{\eta_{line}}
\gtrsim \left\{\begin{array}{cc}50 L_{line}, & $ Fe line$\\
20 L_{line} ,& $ soft narrow lines$\end{array}
\right.\end{equation}
    where L$_{line}$ can be derived from observation of the flux
  of the X-ray emission lines and in this case a solar metallicity
  is assumed. $\eta$$_{line}$ is estimated in reflect mechanism
  (Lazzati et al.2002).

    In Eq(5), T$_{em}$ can be given by the equation:

\begin{equation}
T_{em} = \frac{3c^{3}I}{B_{p}^{2}R^{6}\Omega_{0}^{2}}
            =2.05\times10^{3}{\rm s}(I_{45}B_{p,15}^{-2}P_{0,-3}^{2}R_{6}^{-6})
\end{equation}
Here B$_{p,15}$=B$_{p}$/(10$^{15}$G), and P$_{0,-3}$ is the
initial rotation period in units of millisecond, and I$_{45}$ is
the moment of inertia in units of 10$^{45}$ g cm$^{2}$, and
$\Omega_{0}$ is the initial angular frequency.

  And

\begin{equation}
L_{0}=L_{em,0}
= \frac{I\Omega_{0}^{2}}{2T_{em}}
\simeq1.0\times10^{49} {\rm erg s}^{-1}(B_{p,15}^{2}P_{0,-3}^{-4}R_{6}^{6})
\end{equation}

    During the time interval T$_{c}$$<$T$<$T$_{em}$, one can
  expect a distinctive achromatic bump to appear in the
  lightcurves (Zhang $\&$ M$\acute{e}$sz$\acute{a}$ros 2001).

    Considering the pulsar case, in Eq.(5), to derive T$_{c}$,
  we take T$<$T$_{em}$, L=L$_{0}$=L$_{em,0}$, therefore q=0 in Eq.(3).
  Further we assume the blast wave is adiabatic, k=0. Eq.(4) can be simplified
  as
\begin{equation}
T_{c} = Max(1,\frac{E_{imp}}{L_{0}T_{0}})T_{0}
\end{equation}

    Under the estimation of Eq.(5), considering T as the time of the
  lines appearing, we can get T$\gg$T$_{em}$. L$_{ill}$ can be taken from Eq.(6).
  We derive the values of B$_{p,15}$ , P$_{0,-3}$ and the critical
  time at which the bump begins to appear.All of them can be found in Table 2.

    Since there is no exact value of the GRB explosion energy for
  GRB030227, we have assumed that the $\gamma$-ray energy is
  $E_{\gamma} \sim $5$\times$10$^{50}$ erg(Frail et al.2001).

\section{Discussion}

    If extremely high energy which accounts for the production of
  the X-ray line features comes from a continuous
  postburst outflow from a highly magnetized millisecond
  pulsar or magnetar, early afterglow light curves may be affected
  by this continuous postburst outflow and show a distinct,
  achromatic bump feature. In this case,the afterglow light curves
  flatten after a critical time T$_{c}$ (Eq.(9)) and steepen again
  after a time T$_{em}$ (Eq.(7)). With the flux of X-ray lines and the time
  when the lines begin to appear, we can obtain the upper limit of the
  critical time T$_{c}$  when the bump appears. It should be noted
  that a solar metallicity is assumed.

    We assume the time when Fe line began to appear was
  t$_{s}$ in Table 1.. Given the value of $\eta_{line}$ (the efficiency
  of conversion of the X-ray ionizing continuum into X-ray line
  photons), L$_{ill}$ can
  be estimated by Eq.(6). So the upper limit of critical time T$_{c}$ can
  be obtained from Eq.(9).

    $\eta$$_{line}$ is adopted to be not more than 2$\%$, and from the
  observed flux of the iron line
  F$_{line}$=3.0$\times$10$^{-13}$ erg cm$^{-2}$s$^{-1}$,
  we can obtain L$_{ill}\gtrsim$ 6.0$\times$10$^{46}$ erg s$^{-1}$ and
  T$_{c}$ $\ll$ 1.1 hrs for GRB970508. It is much earlier than the time
  when optical and X-ray afterglow began to be detected (eg.Fruchter \& Pian 1998,
  Piro et al.1999). So it is too early for the distinctive achromatic bump in
  the afterglow of GRB970508 to be detected.

    The most latest time of the bump to appear is GRB020813
  with the critical time T$_{c} \ll $6.4 hrs. It should be noticed
  that there is only one line (S line) with the
  exact value of flux defined, although Si line was also detected.
  For this GRB, $\eta_{line}$$<$2$\%$ is adopted,we obtain the luminosity of S line
  from the observed flux as L$_{ill} \gtrsim$ 1.6$\times$ 10$^{46}$erg
  s$^{-1}$.

    Recently, Urata et al.(2003) claimed that for GRB020813 they detected an early break
  around 0.2 days after the burst using the data of V-band optical afterglow
  observation. They argued that the early break was unlikely to be a jet
  break but was likely to represent the end of an early bump in the lightcurves.
  In addition, Li et al.(2003) reported an early break about 0.14 days after the
  burst based on the R-band data of the
  KAIT telescope. Although the exact time of the early break does not agree with
  each other, they claimed the early break is more likely to be ascribed to an early
  bump emerged in the lightcurve. If these results are true, then it is obvious
  that the existence of the early break is consistent
  with what we have predicted above that an achromatic bump would appear at
  the time T$_{c} \ll $6.4 hrs. For the result of T$_{em}$$\approx$0.2
  days in Urata et al.(2003) paper, we obtained that the
  continuous injection from GRB020813 would be due to a
  pulsar with B$_{p,15}$$\approx$0.68 and P$_{0}$$\approx$1.98 ms.
  T$_{c}$$\approx$0.3 hrs could also be obtained.
  However, more complete analysis of the V-band data by Gorosabel et
  al.(2003)(including the data of Urata et al.(2003)) indicated that there is
  no evidence for the existence of an early bump break. They obtained a
  jet break at the time 0.33$<$t$_{break,V}$$<$0.88 days, in agreement
  with the results by Covino et al.(2003;t$_{break,V}$=0.50 days). If
  the early bump break reported by Li et al.(2003) and Urata et al.(2003) does
  not exist, we can conclude
  that the early bump should appear at the
  time T$_{em}$$<$ 2 hrs (The optical observation
  began at the time of 2hrs after the burst (Fox et al.2002)).

    For GRB011211, we adopt $\eta_{line}$$<$5$\%$(Lazzati et al.2002).
  In this case, a very early bump will appear at the critical time
  T$_{c}$$<$0.5 hours. The early optical afterglow observation was
  obtained 6 hours after the bursts (Bloom $\&$ Berger 2001).
  And X-ray afterglow was observed about 11 hours after the
  burst(Reeves et al.2002). So the bump is too early to be
  detected.

    Same as GRB011211, for GRB030227 we obtain the critical
  time  T$_{c}$$<$0.2 hrs,but it should be noticed that we assume
  the fireball energy is 5$\times$10$^{50}$ erg (Frail et
  al.2001) since we do not know the exact energy of this
  burst. For GRB030227, early optical afterglow observation was
  obtained about 2 hours after the burst (Izumiura et al.2003).
  XMM-Newton began the X-ray afterglow observation 8 hours after
  the burst (Mereghetti et al.2003,Watson et al.2003). So the bump
  also can not be detected.

    As for GRB970828 and GRB991216,the early bump should appear at
  the time T$_{c}$$<$4.1 hrs and 3.5 hrs respectively. For
  GRB991216, the optical afterglow observation began
  at about 10.8 hrs after the GRB (Halpern et al.2000), and X-ray afterglow
  observation is even
  later (Piro et al,1999). For GRB970828, X-ray afterglow observation began
  about 1.17 day after the
  GRB (Yoshida et al.2001) and optical afterglow observation began at 4 hours
  after the burst (Groot et al.1998). For these two GRBs the observation
  time is also too late to observe the bump.

    However, a slightly weaker luminosity could also explain the Fe
  line or soft narrow lines if assuming a larger metal
  abundance. In this case the upper limit of the critical time will
  be slightly late accordingly, but it is also early enough to evade
  the detection of the achromatic bump in the GRBs.

\section{Conclusions}

    The line emission features observed in the X-ray afterglow of
  GRBs are very luminous, and pose strong limit on the energy which
  produces the line feature. These limits are almost unaffected by
  the abundance of metals and collimation of the illuminating
  continuous. If the energy of GRB explosion contributes to the
  production of the line emission, it leads to much larger
  densities of the material surrounding the
  bursts (Ghisellini et al.2002),which is not consistent with the
  results from broadband spectral fitting
  (Panaitescu $\&$ Kumar 2001).

    We consider that the high energy in the illuminating continuum
  comes from the continuous injection outflow after the GRB
  explosion. We argue that the constraint on the energy can be
  avoided.

    At the early time of the GRB afterglow, this continuous
  injection may contribute to the flux of the afterglow, and it
  will produce a distinct achromatic bump in the light
  curves. Focusing on a highly magnetized millisecond pulsar or magnetar, we
  have obtained the critical time when this bump began to appear. For the
  GRBs discussed above, we find that the bump will appear
  in the very early afterglow. We hope this bump may be observed
  by Swift.

    In conclusion, we have presented a scenario in which the bump
  can appear in the early afterglow light curves if we consider
  a continuous postburst outflow comes from a highly
  magnetized millisecond pulsar. This scenario can be tested by the observations
  of Swift satellite in near future.

\acknowledgments We are very grateful to the referee for several
important comments that improved this paper.This work is supported
by the National Natural Science Foundation (grants
10073022,10233010 and 10225314) and the National 973 Project on
Fundamental Researches of China (NKBRSF G19990754).

\begin{deluxetable}{crrrrrrrrrrrr}
\tabletypesize{\scriptsize} \tablecaption{Properties of the X-ray emission lines
detected so far. \label{tbl-1}} \tablewidth{0pt}
\tablehead{\colhead{GRB} & \colhead{Z$_{opt}$}  & \colhead{F$^{line}_{-14}$} &
\colhead{$\epsilon^{line}$} & \colhead{ t$_{s}$-t$_{e}$} &
\colhead{L$^{iso}_{44}$} & \colhead{E$^{iso}_{49}$} &
\colhead{$\theta_{j}$} & \colhead{n}  &
\colhead{E$^{F01}_{\gamma,50}\tablenotemark{c}$}   &
\colhead{E$^{B03}_{\gamma,50}\tablenotemark{d}$} &
\colhead{E$_{\gamma,50}\tablenotemark{e}$ }&
\colhead{Ref.}
}\startdata
970508 & 0.835 & 30$\pm$10  & 3.4$\pm$0.3     & 6-16   & 12$\pm$4    & 2.25-3.6
                & 16.7 & 1.00 & 2.34 & 3.84 & 110-180$\tablenotemark{f}$ &1,8,9,10  \\
970828 & 0.958 & 15$\pm$8   & 5$\pm$0.25      & 32-38  & 8.1$\pm$4.3 & 0.9-5
                & 4.1  &      & 5.75 & 17.00& 45-250  &2,8,9 \\
991216 & 1.02  & 17$\pm$5   & 3.5$\pm$0.06    & 37-40  & 11$\pm$3    & 0.6-7.7
                & 2.9  & 4.70 & 6.95 & 17.07& 30-380  &3,8,9,11 \\
000214 & 0.46$\tablenotemark{a}$& 6.7$\pm$2.2& 4.7$\pm$0.2& 12-41  & 0.6$\pm$0.2 & 0.4-0.6
                &      &      &      &      & 20-30   &4 \\
011211 & 2.14  & 4.0$\pm$1.6& Tot$\tablenotemark{b}$& 11-12.4& 15.6$\pm$6.3& 0.25-2.2
                & 3.6  &      & 1.32 & 4.17 & 5-44   &5,8,9  \\
020813 & 1.254 & 1.6$\pm$0.8&       S       & 21-42.7& 1.7$\pm$0.85& 0.6-1.2
                & 1.8  &      & 3.66 & 11.58& 30-60   &6,8,9 \\
030227 & 1.6   & 17.8     & Tot & 19.4-22& 34.4      & 1.2-10.5
                &      &      &      &      & 24-210  &7 \\
\enddata

\tablenotetext{a}{this burst does not have an optical
determination of the redshift,which is inferred from the X-ray
line,identified as a iron K$_{\alpha}$ line.}
\tablenotetext{b}{Tot means $\epsilon^{line}$ is the total energy
                  of soft lines including S,Si,Ar,Mg and Ca. }
\tablenotetext{d}{total energy of the GRB prompt explosion
     in units of 10$^{50}$ erg
    corrected by the degree of the collimation by Frail et
    al.(2001)}
\tablenotetext{d}{total energy of the GRB prompt explosion
       in units of 10$^{50}$ erg
      corrected by the degree of the collimation by  Bloom et
      al.(2003)}
\tablenotetext{e}{total energy in $\gamma$-rays in units of
10$^{50}$ erg as measured from the emission lines fluence.}
\tablenotetext{f}{Two values in this column are corresponding to the
short lived line and long lived line respectively}

\tablecomments{The units of F$^{line}_{-14}$,$\epsilon^{line}$,
t$_{s}$-t$_{e}$,L$^{iso}_{44}$,E$^{iso}_{49}$,$\theta_{j}$ and n
are erg cm$^{-2}$ s$^{-1}$,keV,hrs,erg s$^{-1}$,erg,degree and cm$^{-3}$.
 We use the notation
 Q$\equiv$$10^{x}Q_{x}$.References:1:Piro et al.1999;2:Yoshida et al.
 3:Piro et al.2000;4:Antonelli et al.2000;5:Reeves et al.2002;6:Butle
 7:Watson et al.2003;8:Frail et al.2001;9:Bloom et al.2003;10:Frail e
 11:Panaitescu et al.2001.}

\end{deluxetable}

\clearpage

\begin{table}
\begin{center}
\caption{Properties of the highly magnetized millisecond pulsar
and the critical time of the achromatic bump appearing.\label{tbl-2}}
\begin{tabular}{crrrr}
\tableline\tableline

         GRB & L$_{ill,46}$ & B$_{p,15}$ & P$_{0,-3}$ & T$_{c}$(h)\\
\tableline
         GRB970508 & 6.0 & 1.2  & $<3.97$ & $<1.1$ \\
         GRB970828 & 4.0 & 0.23 & $<2.12$ & $<4.1$ \\
         GRB991216 & 5.5 & 0.21 & $<1.67$ & $<3.5$ \\
         GRB011211 & 7.8 & 0.58 & $<2.56$ & $<0.5$ \\
         GRB020813\tablenotemark{a} & 1.6  & 0.68 & $<4.12$ & $<6.4$ \\
         GRB030227\tablenotemark{b} & 68.8 & 0.11 & $<0.65$ & $<0.2$ \\
\tableline
\end{tabular}
\tablenotetext{a}{The estimation of the luminosity of the ionizing
continuum which produce the emission line feature in this burst is
from S line although Si line is also detected.$\eta_{line}$ is
taken to be 2$\%$.}
\tablenotetext{b}{To estimate T$_{c}$ we assume the fireball energy
of GRB is 5$\times$10$^{50}$ erg(Frail et al.2001). }
\end{center}
\end{table}

\end{document}